\documentclass{jfm}
\usepackage{graphicx,bm}
\def\dt{{\rm d}t}
\def\d{{\rm d}}

\def\dk{{\rm d}k}

\def\x{{\bm x}}
\def\k{{\bm k}}
\def\etal{{\it et al. }}

\title[Energy spectra in SQG turbulence]
{Large-scale energy spectra in surface quasi-geostrophic turbulence}
\author[C.V. Tran and J.C. Bowman]{C\ls H\ls U\ls O\ls N\ls G\ns V.\ns 
T\ls R\ls A\ls N\thanks{Present address: Mathematics Institute,
University of Warwick, Coventry CV4 7AL, UK}\ns \and J\ls O\ls H\ls 
N\ns C.\ns B\ls O\ls W\ls M\ls A\ls N\ns}
\affiliation{Department of Mathematical and Statistical Sciences,\\
University of Alberta, Edmonton, Alberta, Canada, T6G 2G1}

\date{12 May 2004 and in revised form 05 November 2004}

\begin{document}

\maketitle

\begin{abstract}
The large-scale energy spectrum in two-dimensional turbulence governed 
by the surface quasi-geostrophic (SQG) equation 
$$\partial_t(-\Delta)^{1/2}\psi+J(\psi,(-\Delta)^{1/2}\psi)
=\mu\Delta\psi+f$$
is studied. The nonlinear transfer of this system conserves the two 
quadratic quantities $\Psi_1=\langle[(-\Delta)^{1/4}\psi]^2\rangle/2$ 
and $\Psi_2=\langle[(-\Delta)^{1/2}\psi]^2\rangle/2$ (kinetic energy), 
where $\langle\cdot\rangle$ denotes a spatial average. The energy 
density $\Psi_2$ is bounded and its spectrum $\Psi_2(k)$ is shallower 
than $k^{-1}$ in the inverse-transfer range. For bounded turbulence, 
$\Psi_2(k)$ in the low-wavenumber region can be bounded by $Ck$ where 
$C$ is a constant independent of $k$ but dependent on the domain size. 
Results from numerical simulations confirming the theoretical 
predictions are presented. 
\end{abstract}

\section{Introduction}

The dynamics of a three-dimensional stratified rapidly rotating fluid 
is characterized by the geostrophic balance between the Coriolis force 
and pressure gradient. The nonlinear dynamics governed by the first-order
departure from this linear balance is known as quasi-geostrophic
dynamics and is inherently three-dimensional. The theory of 
quasi-geostrophy is interesting and the research performed on this subject 
constitutes a rich literature (see, for example, Charney 1948, 1971; 
Rhines 1979; Pedlosky 1987). This theory renders a variety of 
two-dimensional models that are appealing for their relative simplicity 
and yet sufficiently sophisticated to capture the underlying dynamics 
of geophysical fluids. One such model, the so-called surface 
quasi-geostrophic (SQG) equation, is the subject of the present study.

Quasi-geostrophic flows can be described in terms of the 
geostrophic streamfunction $\psi(\x,t)$. The 
vertical dimension $z$ is usually taken to be semi-infinite and the 
horizontal extent may be either bounded or unbounded. Normally, decay 
conditions are imposed as $z\rightarrow\infty$. At the flat surface 
boundary $z=0$, the vertical gradient of $\psi(\x,t)$ matches the 
temperature field $T(\x,t)$, 
i.e. $T(\x,t)|_{z=0}=\partial_z\psi(\x,t)|_{z=0}$. 
For flows with zero potential vorticity, this surface temperature 
field can be identified with $(-\Delta)^{1/2}\psi$, where $\Delta$ is 
the (horizontal) two-dimensional Laplacian. Here, the operator 
$(-\Delta)^{1/2}$ is defined by 
$(-\Delta)^{1/2}\widehat\psi(\k)=k\widehat\psi(\k)$, where $k=|\k|$ is 
the wavenumber and $\widehat\psi(\k)$ is the Fourier transform of 
$\psi(\x)$. The conservation equation 
governing the advection of the temperature $(-\Delta)^{1/2}\psi$ by 
the surface flow is (Blumen 1978; Pedlosky 1987; 
Pierrehumbert, Held \& Swanson 1994; Held \etal 1995) 
\begin{eqnarray}
\label{Tadvection}
\partial_t(-\Delta)^{1/2}\psi+J(\psi,(-\Delta)^{1/2}\psi)&=&0,
\end{eqnarray}
where $J(\varphi,\phi)=\partial_x\varphi\partial_y\phi
-\partial_y\varphi\partial_x\phi$.
This equation is known as the SQG equation.

In this paper a forced-dissipative version of (\ref{Tadvection}) 
is studied. A dissipative term of the form 
$\mu\Delta\psi$, where $\mu>0$, which results from Ekman pumping 
at the surface, is considered (Constantin 2002; Tran 2004). Since 
$(-\Delta)^{1/2}\psi$ is the advected quantity, this physical 
dissipation mechanism corresponds to the (hypoviscous) dissipation 
operator $\mu(-\Delta)^{1/2}$. The dissipation coefficient $\mu$ has 
the dimension of velocity and is not vanishingly small in the 
atmospheric context (Constantin 2002). The system is assumed 
to be driven by a forcing $f$, for which the spectral support is 
confined to wavenumbers $k\ge s>0$ (in bounded turbulence, wavenumber
zero is replaced by the minimum wavenumber). 
Thus, the forced-dissipative SQG equation can be written as 
\begin{eqnarray}
\label{governing}
\partial_t(-\Delta)^{1/2}\psi+J(\psi,(-\Delta)^{1/2}\psi)
&=&\mu\Delta\psi+f.
\end{eqnarray}
It is customary in the classical theory of turbulence to consider a
doubly periodic domain of size $L$; the unbounded case is obtained 
{\it via} the limit $L\rightarrow\infty$.

The Jacobian operator $J(\cdot,\cdot)$ admits the identities
\begin{eqnarray}
\label{id}
\langle\chi J(\varphi,\phi)\rangle=-\langle\varphi J(\chi,\phi)\rangle
=-\langle\phi J(\varphi,\chi)\rangle,
\end{eqnarray}
where $\langle\cdot\rangle$ denotes the spatial average. As a consequence, 
the nonlinear term in (\ref{governing}) obeys the conservation laws
\begin{eqnarray}
\label{conservation}
\langle\psi J(\psi,(-\Delta)^{1/2}\psi)\rangle=
\langle(-\Delta)^{1/2}\psi J(\psi,(-\Delta)^{1/2}\psi)\rangle=0.
\end{eqnarray}
It follows that the two quadratic quantities $\Psi_\theta=\langle
|(-\Delta)^{\theta/4}\psi|^2\rangle/2=\int\Psi_\theta(k)\,\dk$,
where $\theta=1,2$, are conserved by nonlinear transfer. Here, 
$\Psi_\theta(k)$ is defined by $\Psi_\theta(k)=k^\theta\Psi(k)$, 
$\Psi(k)$ is the power density of $\psi$ associated with wavenumber 
$k$ and $\theta$ is a real number. Note that $\Psi_2(k)$ is the kinetic 
energy spectrum and $\Psi_2$ is the kinetic energy density.

The simultaneous conservation of two quadratic quantities by advective 
nonlinearities is a common feature in incompressible fluid systems in 
two dimensions. Some familiar systems in this category are the 
Charney--Hasegawa--Mima equation (Hasegawa \& Mima 1978; Hasegawa, 
Maclennan \& Kodama 1979) and the class of $\alpha$ turbulence equations 
(Pierrehumbert \etal 1994), which includes both the Navier--Stokes
and the SQG equations. These conservation laws, together with the 
scale-selectivity of the dissipation and unboundedness of the domain, 
are the building block of the classical dual-cascade theory 
(Fj{\o}rtoft 1953; Kraichnan 1967, 1971; Leith 1968; Batchelor 1969). 
This theory, when applied to the present case, implies that $\Psi_1$ 
cascades to low wavenumbers (inverse cascade) and $\Psi_2$ cascades 
to high wavenumbers (direct cascade). For some recent discussion on 
the possibility of a dual cascade in various two-dimensional systems, 
including the Navier--Stokes and SQG equations, see \cite{TS02}, 
Tran \& Bowman (2003b,2004) and \cite{T04}. The inverse
cascade toward wavenumber $k=0$ would eventually evade viscous 
dissipation altogether because the spectral dissipation rate vanishes 
as $k\rightarrow0$. Hence, according to the classical picture, 
$\Psi_1$ necessarily grows unbounded, by a steady growth rate 
$\d\Psi_1/\dt>0$, as $t\rightarrow\infty$.
Strictly speaking, one may have to address the possibility of a 
dissipated inverse cascade, 
i.e. one for which the dissipation of $\Psi_1$ occurs at scales much 
larger than the forcing scale and for which $\d\Psi_1/\dt$ has a zero time
mean. Such a cascade is not a plausible scenario (and is not the 
traditional undissipated inverse cascade) in fluid systems, dissipated 
by a single viscous operator, where the viscous dissipation rate diminishes 
toward the large scales. A discussion of this issue can be found 
in \cite{T04}.
  
In this study, upper bounds are derived for the time averages of the 
kinetic energy density $\Psi_2$ and of the large-scale spectrum 
$\Psi_2(k)$. These bounds are derived from the governing equation, 
involving simple but rigorous estimates. The bound on $\Psi_2$ is valid 
in both unbounded and bounded cases, and a straightforward consequence
of this bound is a bound on the energy spectrum, which also applies to 
both unbounded and bounded turbulence. Another bound on the large-scale 
energy spectrum is derived by estimating the nonlinear triple-product
term representing the inverse transfer of $\Psi_1$. This result applies 
to bounded turbulence since upper bounds for the triple-product term 
are inherently domain-size dependent. The difficulties of extending this 
result to the unbounded case are discussed. Some numerical results 
confirming the theoretical predictions are presented.

\section{Bounded dynamical quantities}

A notable feature of unbounded incompressible fluid turbulence in two
dimensions is the appearance of infinite quadratic quantities (per unit
area): namely, the kinetic energy density $\Psi_2$ for Navier--Stokes
turbulence and $\Psi_1$ for the SQG case. According to the classical theory
(applied to the SQG case), a (steady) injection of $\Psi_1$, applied around
some finite wavenumber $s$, cascades to ever-larger scales, leading to an
unbounded growth of $\Psi_1$ (this is presumably the case for the general
quadratic invariant $\Psi_\alpha$ in the so-called $\alpha$ turbulence;
{\it cf.} Tran 2004).  In other words, if the classical inverse cascade is
realizable, unbounded incompressible fluid turbulence in two dimensions
constitutes an ill-posed problem, in the sense that a key quadratic
invariant becomes infinite. Of course, there still exist finite quadratic
quantities, in particular the dissipation agent for each quadratic
invariant. This section is concerned with these quantities.

On multiplying (\ref{governing}) by $\psi$ and $(-\Delta)^{1/2}\psi$ 
and taking the spatial average of the resulting equations, noting from 
the conservation laws that the nonlinear terms identically vanish, one 
obtains evolution equations for $\Psi_1$ and $\Psi_2$,
\begin{eqnarray}
\label{Psi1evolution}
\frac{\d}{\dt}\Psi_1&=&-2\mu\Psi_2+\langle f\psi\rangle,\\
\label{Psi2evolution}
\frac{\d}{\dt}\Psi_2&=&-2\mu\Psi_3
+\langle f(-\Delta)^{1/2}\psi\rangle.
\end{eqnarray}
Using the Cauchy--Schwarz and geometric--arithmetic mean inequalities, one
obtains upper bounds on the injection terms in (\ref{Psi1evolution}) and
(\ref{Psi2evolution}):
\begin{eqnarray}
\label{forcebounds}
\langle f\psi\rangle&\le&\langle|(-\Delta)^{1/2}\psi|^2\rangle^{1/2}
\langle|(-\Delta)^{-1/2}f|^2\rangle^{1/2}
\le\mu\Psi_2+\mu^{-1}F_{-2},\nonumber\\
\langle f(-\Delta)^{1/2}\psi\rangle &\le&
\langle|(-\Delta)^{3/4}\psi|^2\rangle^{1/2}
\langle|(-\Delta)^{-1/4}f|^2\rangle^{1/2}
\le\mu\Psi_3+\mu^{-1}F_{-1},
\end{eqnarray}
where the `integration by parts' rule $\langle(-\Delta)^\theta\phi
\chi\rangle=\langle(-\Delta)^{\theta'}\phi(-\Delta)^{\theta''}\chi\rangle$,
for $\theta=\theta'+\theta''$, has been used and $F_\theta=\langle
|(-\Delta)^{\theta/4}f|^2\rangle/2$.  
Substituting (\ref{forcebounds}) in (\ref{Psi1evolution}) and 
(\ref{Psi2evolution}) yields
\begin{eqnarray}
\label{evolbound1}
\frac{\d}{\dt}\Psi_1&\le&-\mu\Psi_2+\mu^{-1}F_{-2},\\
\label{evolbound2}
\frac{\d}{\dt}\Psi_2&\le&-\mu\Psi_3+\mu^{-1}F_{-1}.
\end{eqnarray}
To avoid unnecessary complications, zero initial conditions are assumed,
so that for $T>0$ the time means $\langle\d\Psi_1/\dt\rangle_t=
\Psi_1(T)/T$ and $\langle\d\Psi_2/\dt\rangle_t$ are non-negative. 
%The time-asymptotic mean $\left\langle \d\Psi_1/\d t
%\right\rangle_t=\lim_{T\rightarrow \infty} \frac{1}{T}\int_0^T \d\Psi_1/\d t
% \ge 0$ (whenever it exists) since
%$\int_0^T \d\Psi_1/\d t\ge-\Psi_1(0)$ for any $T$. Similarly, $\left\langle
%\d\Psi_2/\d t \right\rangle_t \ge 0$.
One can then deduce upper bounds on the time means 
$\langle\Psi_2\rangle_t$ and $\langle\Psi_3\rangle_t$, which are valid 
regardless of whether or not $\Psi_1$ remains finite in the limit 
$t\rightarrow\infty$:
\begin{eqnarray}
\label{averagebound1}
\langle\Psi_2\rangle_t &\le& \mu^{-2}\langle F_{-2}\rangle_t,\\
\label{averagebound2}
\langle\Psi_3\rangle_t &\le& \mu^{-2}\langle F_{-1}\rangle_t.
\end{eqnarray}

For $\theta\in(2,3)$, $\langle\Psi_\theta\rangle_t$ is also bounded.
Indeed, from the H\"{o}lder inequalities 
$\Psi_\theta\le\Psi_2^{3-\theta}\Psi_3^{\theta-2}$ ({\it cf.} Tran 2004) and 
$\langle\Psi_2^{3-\theta}\Psi_3^{\theta-2}\rangle_t\le
\langle\Psi_2\rangle_t^{3-\theta}\langle\Psi_3\rangle_t^{\theta-2}$,
one can deduce from (\ref{averagebound1}) and (\ref{averagebound2}) that
\begin{eqnarray}
\langle\Psi_\theta\rangle_t &\le& 
\langle\Psi_2\rangle_t^{3-\theta}\langle\Psi_3\rangle_t^{\theta-2}
\le \mu^{-2}
\langle F_{-2}\rangle_t^{3-\theta}\langle F_{-1}\rangle_t^{\theta-2}.
\end{eqnarray}
This result implies that for $\theta\in(2,3)$, 
$\langle\Psi_\theta\rangle_t$ is bounded, provided that both 
$\langle F_{-1}\rangle_t$ and $\langle F_{-2}\rangle_t$ are bounded. 
This condition is assured if $s>0$ and $F_0$ is bounded, a condition 
normally required of the forcing, because $F_{-2}\le F_{-1}/s\le F_0/s^2$. 
One may even consider a class of forcing for which $F_0=\infty$ and
$F_{-2}\le F_{-1}/s<\infty$. 

Upper bounds of the above type on dynamical quantities are rather 
trivial for bounded turbulence. However, they are important in the 
unbounded case, for two reasons. First, the scale-selective viscous 
dissipation allows for the possibility of unbounded growth of certain 
quadratic quantities toward the low wavenumbers. Hence, rigorous
bounds on dynamical quantities are not as abundant as in the bounded
case. Second, analytic studies of the nonlinear triple-product transfer
function are difficult in unbounded domains. In the absence of pointwise 
estimates for the spectrum, these bounds are particularly useful for 
qualitative estimates of the large-scale distribution of energy. For 
example, \cite{T04} uses inequality (\ref{averagebound1}) to argue that 
the energy spectrum $\Psi_2(k)$ should be shallower than $k^{-1}$,
as $k\rightarrow0$. 

\section{Large-scale energy spectrum}

In this section, it is shown that the physical laws of SQG dynamics 
admit only large-scale energy spectra shallower than $k^{-1}$. This 
result is due in part to the fact that the simultaneous conservation
of $\Psi_1$ and $\Psi_2$ allows virtually no kinetic energy to get
transferred toward the low wavenumbers, so that only large-scale kinetic 
energy spectra shallower than $k^{-1}$ are possible.

\subsection{Shell-averaged energy spectrum}
For a given wavenumber $r$, let us denote by $S=S(r)$ the wavenumber shell
between $k=r/2$ and $k=3r/2$, i.e. $S(r)=\{\k : r/2 \le k \le 3r/2\}$.
The shell-averaged energy spectrum $\overline\Psi_2(r)$ over $S(r)$ is
defined by
\begin{eqnarray}
\label{spectrum}
\overline\Psi_2(r)&=&\frac{1}{r}\int_{r/2}^{3r/2}\Psi_2(k)\,\dk.
\end{eqnarray}
In the present case of a doubly periodic domain of size $L$, the Fourier 
representation of the stream function is 
$\psi(\x)=\sum_{\k}\exp\{i\k\cdot\x\}\widehat\psi(\k)$, where 
$\k=2\pi L^{-1}(k_x,k_y)$ with $k_x$ and $k_y$ being integers not 
simultaneously zero. Let $\psi(S)$ denote the component of $\psi$ 
spectrally supported by $S$, i.e. $\psi(S)=\sum_{\k\in S}\exp
\{i\k\cdot\x\}\widehat\psi(\k)$. One has 
\begin{eqnarray}
\label{ineq}
\sup_{\x}|\nabla\psi(S)| &\le& \sum_{\k\in S}k|\widehat\psi(\k)| 
\le \left(\sum_{\k\in S}1\sum_{\k\in S}k^2|\widehat\psi(\k)|^2\right)^{1/2}
\le cLr\Psi_2^{1/2}(S),
\end{eqnarray}
where the Cauchy--Schwarz inequality is used, the sum $\sum_{\k\in S}1=(cLr)^2$
is the number of wavevectors in $S$, $c$ is an absolute constant of 
order unity and $\Psi_2(S)$ is the contribution to the kinetic energy 
from $S$.

\subsection{Upper bounds for the energy spectrum}
A simple upper bound for $\overline\Psi_2(k)$, which is applicable to both the
unbounded and bounded cases, can be derived from (\ref{averagebound1}).
In fact, it follows from (\ref{averagebound1}) and (\ref{spectrum}) that
\begin{eqnarray}
\label{spectbound1}
\langle\overline\Psi_2(k)\rangle_t&=&\frac{1}{k}\int_{k/2}^{3k/2}
\langle\Psi_2(\kappa)\rangle_t\,\d\kappa
\le \mu^{-2}\langle F_{-2}\rangle_tk^{-1}.
\end{eqnarray}
This bound is supposed to apply to $k$ in the inverse-transfer region.
For $k$ in the direct-transfer region, (\ref{averagebound2}) yields
\begin{eqnarray}
\label{spectbound2}
\langle\overline\Psi_2(k)\rangle_t&=&\frac{1}{k}\int_{k/2}^{3k/2}
\langle\Psi_2(\kappa)\rangle_t\,\d\kappa
\le \frac{2}{k^2}\int_{k/2}^{3k/2}
\langle\Psi_3(\kappa)\rangle_t\,\d\kappa
\le 2\mu^{-2}\langle F_{-1}\rangle_tk^{-2}.
\end{eqnarray}

The upper bound (\ref{spectbound1}) suggests that dimensional analysis
arguments, which predict a large-scale $k^{-1}$ energy spectrum, are not well 
justified. If a persistent inverse cascade of $\Psi_1$ exists 
($\d \Psi_1/\dt > 0$), then the energy $\Psi_2$ ought to acquire a value
such that $\Psi_2<\mu^{-2}F_{-2}$. In the unbounded case, the large-scale
energy spectrum then needs to be strictly shallower than $k^{-1}$, to
ensure that the dissipation of $\Psi_1$ does not increase without bound as
the inverse cascade proceeds toward $k=0$.
On the other hand, if no inverse cascade of $\Psi_1$ exists, then 
a $k^{-1}$ energy spectrum with limited extent is possible. If viscous 
dissipation mechanisms with degrees higher than that of the 
natural dissipation are considered, then the upper bounds derived above 
are not valid. Nevertheless, diminishing energy transfer towards the lowest
wavenumbers appears to be consistent only with spectra shallower than $k^{-1}$
(for low-wavenumber convergence of the energy integral).
The numerical results reported in~\S\,4 are well suited to this expectation.

An upper bound for the large-scale energy spectrum, based on the 
nonlinear transfer term, can be derived for the bounded case. This 
analysis employs elementary but rigorous estimates of the 
triple-product term. For $3k/2<s$, the evolution of $\Psi_1(S(k))$ 
is governed by
\begin{eqnarray}
\frac{\d}{\dt}\Psi_1(S)&=&-\langle\psi(S)
J(\psi,(-\Delta)^{1/2}\psi)\rangle-2\mu\Psi_2(S) \nonumber\\
&=&\langle(-\Delta)^{1/2}\psi J(\psi,\psi(S))\rangle-2\mu\Psi_2(S)\nonumber\\
&\le&\langle|(-\Delta)^{1/2}\psi||\nabla\psi||\nabla\psi(S)|\rangle
-2\mu\Psi_2(S)\nonumber\\
&\le&\sup_{\x}|\nabla\psi(S)|\langle|(-\Delta)^{1/2}\psi||\nabla\psi|\rangle
-2\mu\Psi_2(S)\nonumber\\
&\le&2cLk\Psi_2^{1/2}(S)\Psi_2-2\mu\Psi_2(S)\nonumber\\
&\le&c^2\mu^{-1}L^2k^2\Psi_2^2-\mu\Psi_2(S)\nonumber\\
&=&c^2\mu^{-1}L^2k^2\Psi_2^2-\mu k\overline\Psi_2(k),
\end{eqnarray}
where the second equality is a consequence of (\ref{id}) and the second last
and last inequalities follow from (\ref{ineq}) and the geometric--arithmetic
mean inequality, respectively.
It follows that
\begin{eqnarray}
\label{spectbound3}
\langle\overline\Psi_2(k)\rangle_t&\le&c^2\mu^{-2}L^2k\langle\Psi_2^2\rangle_t.
\end{eqnarray}
A notable feature of (\ref{spectbound3}) is its dependence on the 
fluid domain size. The presence of $L$ in this upper bound is natural: 
the upper bound $\sup_{\x}|\nabla\psi(S)|$, which is associated with 
the fluid velocity at scales $\approx k^{-1}$, is inherently
domain-size dependent. There are no known analytic estimates that allow
one to derive an upper bound on the nonlinear transfer function
$\langle\psi(S)J(\psi,(-\Delta)^{1/2}\psi)\rangle$ in 
terms of `intensive quantities' only. This difficulty arises not only
in the present estimate but also in other analytic estimates 
of the transfer function. In other words, the nonlinear triple-product 
term is intrinsically domain-size dependent. This problem considerably 
limits our ability to assess the nonlinear transfer in unbounded systems. 
Finally, it is worth mentioning that although the upper bound 
(\ref{spectbound3}) has a linear dependence on $k$, it may be more 
excessive than the bound $\mu^{-2}\langle F_{-2}\rangle_t k^{-1}$ 
derived earlier (even for very low wavenumbers). The reason is that 
$L^2k\ge k^{-1}$ and the prefactor $c^2\langle\Psi_2^2\rangle_t$ 
may not be as optimal as $\langle F_{-2}\rangle_t$.

\section{Numerical results}

This section reports results from numerical simulations that illustrate 
the realization of large-scale spectra shallower than $k^{-1}$.  
Equation (\ref{governing}) is simulated in a doubly periodic square of 
side $2\pi$, where the forcing $\widehat{f}(\bm k)$ is nonzero only for 
those wavevectors $\bm k$ having magnitudes lying in the interval 
$K=[59,61]$:
\begin{eqnarray}
\label{forcing}
\widehat{f}(\bm k)&=&\frac{\epsilon}{N}\frac{\widehat{\psi}(\bm{k})}
{2\Psi_1(k)}.
\end{eqnarray} 
Here $\epsilon=1$ is the constant energy injection rate and~$N$ is the 
number of distinct wavenumbers in~$K$. The (constant) injection rate of 
$\Psi_1$ is $\epsilon/s\approx1/60$, where $1/s\approx1/60$ is the mean 
of $k^{-1}$ over $K$. This type of forcing was used by \cite{Shepherd87}, 
\cite{T04} and \cite{TB04} in numerical simulations of a large-scale 
zonal jet on the so-called beta-plane and of Navier--Stokes turbulence.  
The attractive aspect of (\ref{forcing}), as noted in \cite{Shepherd87}, 
is that it is steady. Dealiased $683^2$ and $1365^2$ pseudospectral 
simulations ($1024^2$ and $2048^2$ total modes) were performed. 
Three dissipative forms were considered: $2.5\times10^{-2}\Delta\psi$, 
$-4\times10^{-4}(-\Delta)^{3/2}\psi$, and 
$-6\times10^{-6}\Delta^2\psi+\mu\Delta\psi$ for several values of $\mu$.
The first case represents the natural dissipation of the SQG dynamics
due to Ekman pumping, as mentioned earlier. The second case represents
thermal (molecular) diffusion since $(-\Delta)^{1/2}\psi$ is equivalent 
to the fluid temperature. The third case---the mixed hyperviscous/Ekman 
dissipation form---is considered in order to demonstrate that even slight 
amounts of Ekman damping will inhibit the formation of an inverse cascade.
Unlike \cite{Smith02}, the case of mechanical friction
[$\propto(-\Delta)^{1/2}\psi$] was not considered. The higher resolution 
was used for the first (natural dissipation) case and the lower resolution 
was used for the second and third cases. All simulations were initialized 
with the spectrum $\Psi_2(k)=10^{-5}\pi k/(60^2+k^2)$. 

\begin{figure}
\centerline{\includegraphics{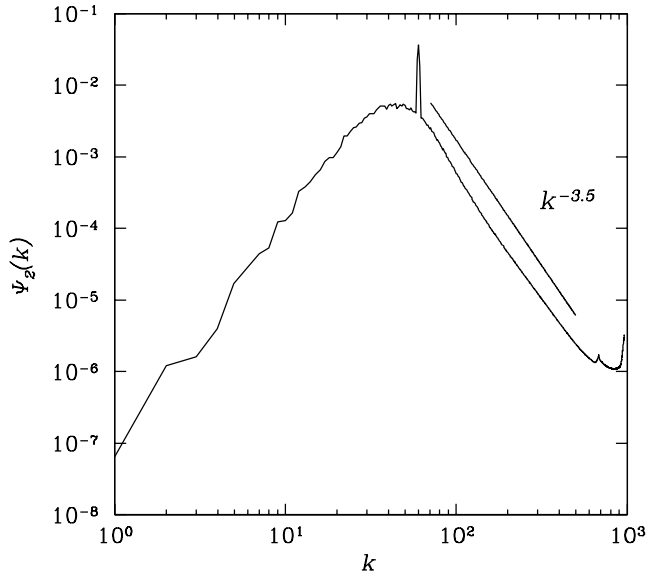}}
\caption{The time-averaged steady-state energy spectrum $\Psi_2(k)$ {\it vs.}\ 
$k$ for the dissipation term $2.5\times10^{-2}\Delta\psi$.}
\label{sqg2}
\end{figure}

Figure~\ref{sqg2} shows the time-averaged steady-state kinetic energy
spectrum for the case of the natural dissipation term
$2.5\times10^{-2}\Delta\psi$. The dissipation agents of $\Psi_1$ and
$\Psi_2$ are, respectively, $\Psi_2$ (energy) and $\Psi_3$. 
The value of the energy, $0.3333$, implies that 
the dissipation of $\Psi_1$, averaged in the same period, is $0.01666$. 
This amounts to virtually all of the injection rate $1/60$. Hence, there
exists no inverse cascade of $\Psi_1$ to the large scales and both
$\Psi_1$ and $\Psi_2$ are steady. The small-scale energy
spectrum scales as $k^{-3.5}$, so that the spectrum $\Psi_3(k)$ scales as
$k^{-2.5}$. This scaling means that the energy dissipation occurs mainly
around the forcing region and is consistent with the bound
(\ref{spectbound2}). 

Unlike Navier--Stokes turbulence, for which the inverse energy cascade
is robust and can be simulated at relatively low resolution, it was
noticed that no choice for the value of $\mu$ at the present resolution
could be used to simulate an inverse cascade of $\Psi_1$. It is not known
whether an inverse cascade of $\Psi_1$ is realizable at higher resolutions,
using a smaller value of $\mu$. Nevertheless, this observation 
suggests that $\Psi_1$ is `reluctant' to cascade to the large scales, 
as compared with the more robust inverse energy cascade in 
Navier--Stokes turbulence.

\begin{figure}
\centerline{\includegraphics{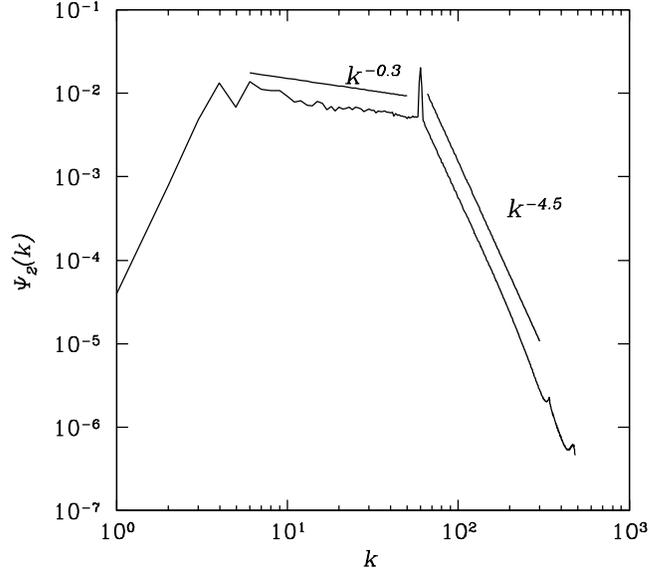}}
\caption{The quasisteady energy spectrum $\Psi_2(k)$ {\it vs.}\ $k$ averaged
between $t=37.3$ and $t=38.7$ for the dissipation term
$-4\times10^{-4}(-\Delta)^{3/2}\psi$.}
\label{sqgviscous1}
\end{figure}

Figure~\ref{sqgviscous1} shows the kinetic energy spectrum averaged
between $t=37.3$ and $t=38.7$, for a lower viscous degree.
The dissipation agents of $\Psi_1$
and $\Psi_2$ are, respectively, $\Psi_3$ and $\Psi_4$ (enstrophy). 
The value of $\Psi_3$ is $20$, implying that the dissipation of $\Psi_1$
is $1.6\times10^{-2}$. This amounts to about $96\%$ of the injection rate
$1/60$. The inverse cascade then carries only a few percent of the
injection of $\Psi_1$ to the large scales.

The small-scale energy spectrum scales as $k^{-4.5}$, so that the 
enstrophy spectrum $\Psi_4(k)$ scales as $k^{-2.5}$. 
Most of the energy dissipation occurs around 
the forcing region, consistent with a `weak' inverse cascade
(one that does not carry virtually all of the injection of $\Psi_1$ toward
$k=0$; {\it cf.} Tran and Bowman 2004, Tran 2004). No direct cascade is
possible for bounded turbulence in equilibrium or for unbounded turbulence
in the presence of a weak inverse cascade. 

\begin{figure}
\centerline{\includegraphics{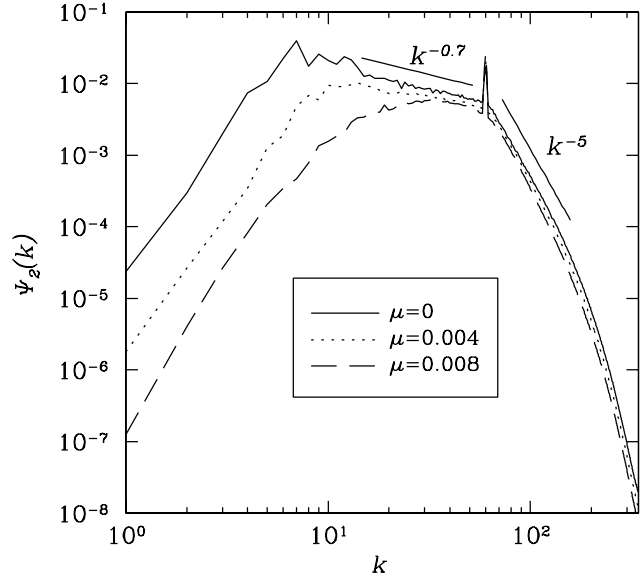}}
\caption{The quasisteady energy spectrum $\Psi_2(k)$ {\it vs.}\ $k$ averaged
between $t=15.67$ and $t=16.52$ for the dissipation term
$-6\times10^{-6}\Delta^2\psi+\mu\Delta\psi$, using three different values of
$\mu$.
} 
\label{sqgviscous2mixed}
\end{figure}

\begin{figure}
\centerline{\includegraphics{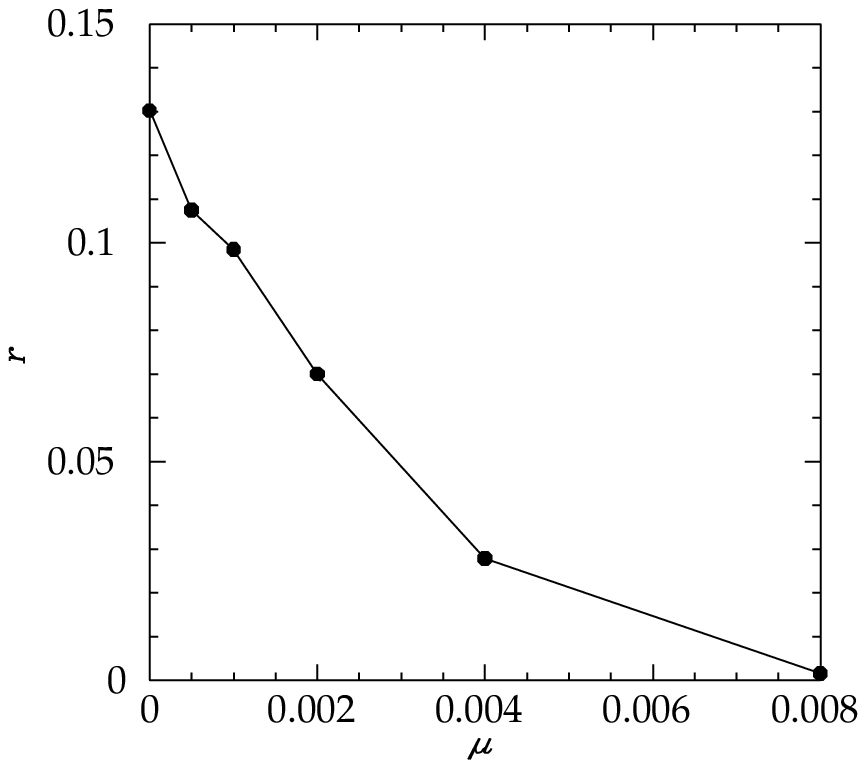}}
\caption{The decay of the inverse cascade strength $r$ for
the dissipation term $-6\times10^{-6}\Delta^2\psi+\mu\Delta\psi$ as $\mu$
is increased.} 
\label{invstrengthvnuL}
\end{figure}

Similarly, Figure~\ref{sqgviscous2mixed} shows the kinetic energy spectrum 
averaged between $t=15.7$ and $t=16.5$ for the mixed dissipation
$-6\times10^{-6}\Delta^2\psi+\mu\Delta\psi$, using three different values of
$\mu$. When $\mu=0$, the dissipation agents of $\Psi_1$ and $\Psi_2$ are,
respectively, $\Psi_4$ (enstrophy) and $\Psi_5$.
The value of the enstrophy, $1208$, implies that the dissipation of $\Psi_1$ is
$1.45\times10^{-2}$, amounting to about $87\%$ of the injection rate $1/60$.
The small-scale energy 
spectrum scales as $k^{-5}$, so that $\Psi_5(k)$ scales as $k^{-2}$. 
Again, this scaling means that most of the energy dissipation occurs around
the forcing region and that the inverse cascade is weak.
We note that as $\mu$ is increased, the inverse cascade becomes
increasingly weak. We emphasize this behaviour by plotting 
in Fig.~\ref{invstrengthvnuL} the inverse cascade {\it strength}
$r=1-2s(\mu\Psi_2+6\times10^{-6}\Psi_4)/\epsilon$ for six different values
of $\mu$.

Unlike Navier--Stokes turbulence, for which the enstrophy acquires its
near-equilibrium value once a discernible inverse-transfer range
has formed, the energy in SQG turbulence can remain significantly less 
than its equilibrium value until a very wide inverse-transfer range 
has developed. For example, for a one-decade Navier--Stokes
inverse-transfer range (achievable in numerical simulations),
the enstrophy acquires 95\% of its projected equilibrium value (calculated
with a $k^{-5/3}$ energy spectrum extrapolated to $k=0$).
On the other hand, for a one-decade SQG inverse-transfer range, the energy
acquires only 66\% of its projected equilibrium value (calculated with a
$k^{-0.7}$ energy spectrum extrapolated to $k=0$, as realized in the 
present simulations; {\it cf.} the $\mu=0$ case of
Figure~\ref{sqgviscous2mixed}). This means
that one needs a considerably wider inverse-transfer region for SQG turbulence 
than for Navier--Stokes turbulence, in order to approach a 
quasi-steady state. This problem is in addition to the resolution 
limitations at the small scales for both cases.  

Due to the steep spectrum in the inverse-transfer region, the energy 
in the $\mu=0$ case of Figure~\ref{sqgviscous2mixed} has not acquired a
value considerably close to its equilibrium value. This means that the
system is still well within the transient phase, However, the dissipation
of $\Psi_1$ (proportional to the enstrophy) cannot grow considerably (without
significant change to the existing  spectrum), because of the high degree
of viscosity, which makes the dissipation of $\Psi_1$ relatively
insensitive to growth of the large-scale energy. 

\section{Conclusion and discussion}

In this paper, the kinetic energy density of SQG turbulence and its 
large-scale spectrum have been studied. For the unbounded case, upper 
bounds are derived for the time means of the kinetic energy density 
and of the large-scale energy spectrum, averaged over a narrow window 
of wavenumbers. Another result is an upper bound on the the time mean 
of the large-scale energy spectrum, which is derived for 
the bounded case. Numerical results confirming the predicted slopes 
of the large-scale energy spectrum are presented and discussed. 

An important feature in SQG turbulence that gives rise to the rigorous 
upper bound on the time mean of the kinetic energy density in the 
unbounded case is that the kinetic energy is the dissipation agent of 
the inverse-cascading candidate $\Psi_1$. This fact is due to the 
hypoviscous nature of the dissipation operator $(-\Delta)^{1/2}$, a 
natural physical dissipation mechanism of SQG dynamics (Ohkitani 1997; 
Constantin 2002; Tran 2004). If $(-\Delta)^{1/2}$ is replaced by an 
operator of the form $(-\Delta)^{\eta}$, where $\eta>1/2$, then the 
simple analysis of Section 2 fails to show that the time mean of the 
energy density $\langle\Psi_2\rangle_t$ is bounded, although it may 
remain so for low degrees of viscosity $\eta$. The reason is that  
the amount of energy getting transferred to wavenumbers lower than a 
given wavenumber $k$ decreases at least as rapidly as $k$, 
so that the spectral dissipation rate $\propto k^{2\eta}$, a consequence
of the dissipation operator $(-\Delta)^{\eta}$, may be  
sufficiently strong to balance the diminishing inverse energy transfer
and keep the energy from growing unbounded. 

Numerical simulations of SQG turbulence were performed, using the 
natural dissipation operator $(-\Delta)^{1/2}$ and two viscous 
operators $\Delta$ and $(-\Delta)^{3/2}$. The results show 
large-scale energy spectra shallower than $k^{-1}$, consistent with the
theoretical prediction.
 
There have been attempts to explain, within the context of SQG 
turbulence (Constantin 2002; Tung \& Orlando 2003), the kinetic energy 
spectra observed in the laboratory experiment of \cite{Baroud02} and 
in the atmosphere. In the former case, the 
turbulence in a rotating tank is driven at a sufficiently small
scale to allow for a wide inverse-transferring range. A $k^{-2}$
spectrum extending over nearly two wavenumber decades lower than
the forcing wavenumber is observed. In the latter case, a 
$k^{-5/3}$ spectrum is observed in the mesoscales (see Frisch 
1995 and Tung \& Orlando 2003 and references therein), which
correspond to wavenumbers higher (lower) than the forcing 
wavenumber if the energy released from baroclinic instability 
(thunderstorms) is considered to be the driving force. The $-2$ power-law
scaling observed in \cite{Baroud02} for the wavenumber range lower than the
forcing wavenumber is excessively steeper than the permissible scalings
derived in this work. The $-5/3$ slope in the atmosphere is either steeper
(if considered to be on the wavenumber range lower than the forcing 
wavenumber) or shallower (if considered to be on the wavenumber range 
higher than the forcing wavenumber) than the permissible slopes.
According to the present analysis, these data cannot be attributed to
SQG turbulence. 

\begin{acknowledgments}
We would like to thank two anonymous referees for their comments,
which were helpful in improving this manuscript. This work was funded by a
Pacific Institute for the Mathematical Sciences Postdoctoral Fellowship, an
Alexander von Humboldt Research Fellowship, and the Natural Sciences and
Engineering Research Council of Canada.
\end{acknowledgments}

%\bibliography{ref}

\end{document}